\documentclass{aastex}
\usepackage{spr-astr-addons}
\usepackage{url}

\shortauthors{Vavrycuk}

\usepackage{graphicx}

\usepackage{natbib}
\usepackage{threeparttable}
\begin{document} 

\title{Impact of galactic and intergalactic dust on the stellar EBL}

\author{V. Vavry\v cuk}

\affil{Institute of Geophysics, The Czech Academy of Sciences}
\affil{Bo\v cn\' i II 1401, 141 00 Praha 4}
\email{vv@ig.cas.cz}

\received{December 17, 2015}

\begin{abstract}
Current theories assume that the low intensity of the stellar extragalactic background light (stellar EBL) is caused by finite age of the Universe because the finite-age factor limits the number of photons that have been pumped into the space by galaxies and thus the sky is dark in the night. We oppose this opinion and show that two main factors are responsible for the extremely low intensity of the observed stellar EBL. The first factor is a low mean surface brightness of galaxies, which causes a low luminosity density in the local Universe. The second factor is light extinction due to absorption by galactic and intergalactic dust. Dust produces a partial opacity of galaxies and of the Universe. The galactic opacity reduces the intensity of light from more distant background galaxies obscured by foreground galaxies. The inclination-averaged values of the effective extinction $A_V$ for light passing through a galaxy is about 0.2 mag. This causes that distant background galaxies become apparently faint and do not contribute to the EBL significantly. In addition, light of distant galaxies is dimmed due to absorption by intergalactic dust. Even a minute intergalactic opacity of $1 \times 10^{-2}$ mag per Gpc is high enough to produce significant effects on the EBL. As a consequence, the  EBL is comparable with or lower than the mean surface brightness of galaxies. Comparing both extinction effects, the impact of the intergalactic opacity on the EBL is more significant than the obscuration of distant galaxies by partially opaque foreground galaxies by factor of 10 or more. The absorbed starlight heats up the galactic and intergalactic dust and is further re-radiated at IR, FIR and micro-wave spectrum. Assuming static infinite universe with no galactic or intergalactic dust, the stellar EBL should be as high as the surface brightness of stars. However, if dust is considered, the predicted stellar EBL is about $290 \, \mathrm{n W m}^{-2}\mathrm{sr}^{-1}$, which is only 5 times higher than the observed value. Hence, the presence of dust has higher impact on the EBL than currently assumed. In the expanding universe, the calculated value of the EBL is further decreased, because the obscuration effect and intergalactic absorption become more pronounced at high redshifts when the matter was concentrated at smaller volume than at present.
\end{abstract}


\section{Introduction}

\begin{figure*}
\centering
\includegraphics[angle=0,width=16cm,trim = {40 65 10 34},clip]{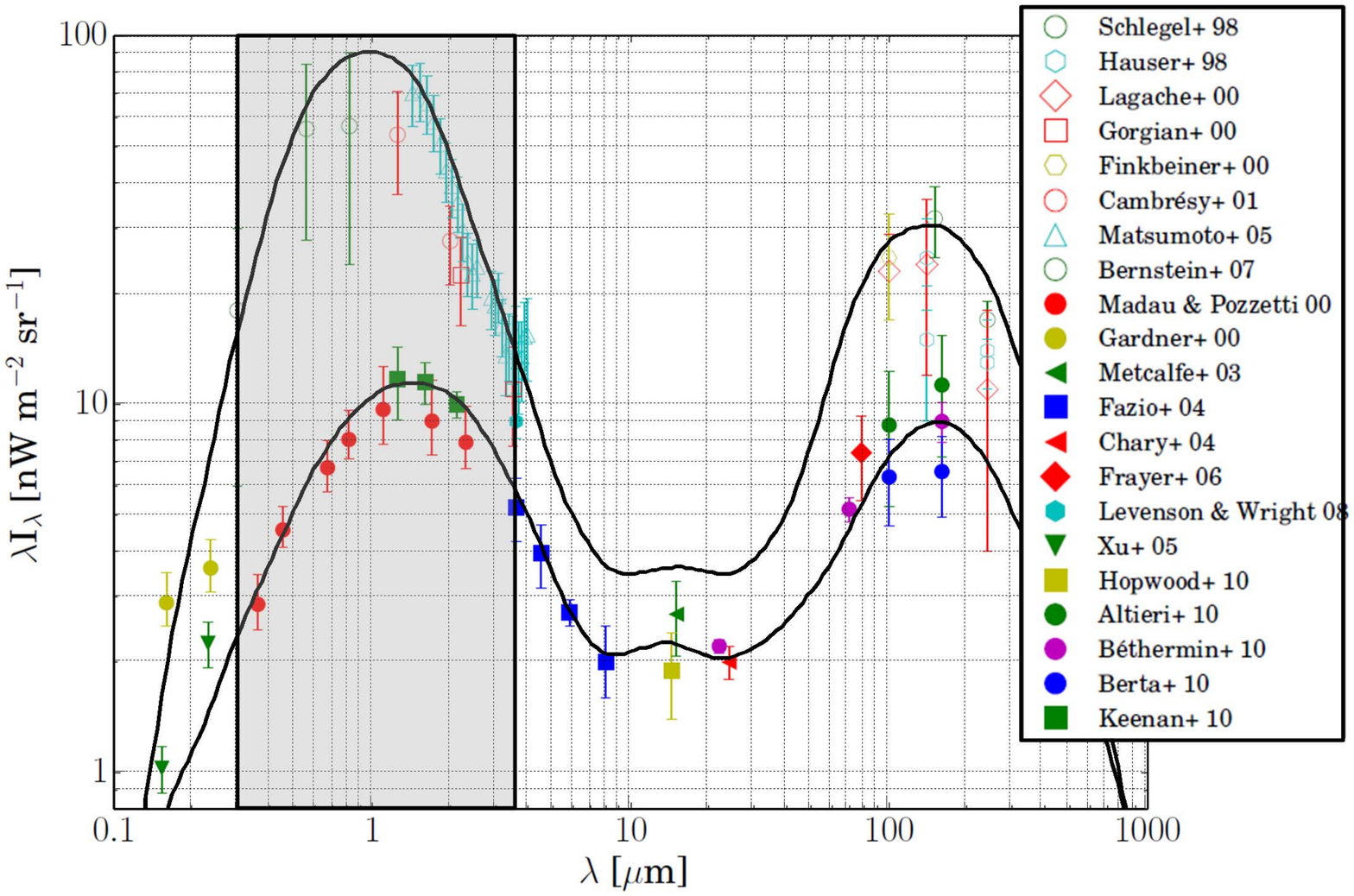}
\caption{
Spectral energy distribution (SED) of the EBL with estimates of its minimum and maximum limits (black lines). The shaded area marks the range of the stellar wavelengths: 300-3500 nm. The observations reported by various authors are marked by colour symbols (modified after \cite{Dominguez2011}).
}
\label{fig:1}
\end{figure*}

The extragalactic background light (EBL) covers the near-ultraviolet, visible and infrared wavelengths from 0.1 to 1000 $\mu$m. Measurements of the EBL are provided by data from the Cosmic Background Explorer (COBE) mission, by the Infrared Space Observatory (ISO) instruments and by the Submillimeter Common User Bolometer Array (SCUBA) instrument (for reviews, see \cite{Hauser2001, Lagache2005}). The direct measurements are supplemented by analysing integrated light from extragalactic source counts which provide a lower limit on the EBL \citep{Madau2000, Hauser2001}. The upper limits are provided by attenuation of gamma rays from distant blazars due to scattering on the EBL \citep{Kneiske2004, Dwek2005, Primack2011, Gilmore2012, Biteau2015}. The spectral energy distribution of the EBL has two distinct maxima: at visible-to-near-infrared wavelengths in the range from 0.7 to 2 $\mu$m associated with the radiation of stars, and at far-infrared wavelengths from 100 to 200 $\mu$m associated with the thermal radiation of cold and warm dust in galaxies \citep{Schlegel1998, Calzetti2000}. Despite the extensive number of measurements of the EBL, the uncertainties in their peak values are still large (see Fig. 1). The total EBL from 0.1 to 1000 $\mu$m lies roughly in the range from 40 to 200  $\mathrm{n W m}^{-2}\mathrm{sr}^{-1}$ and half of this value comes from the visible-to-near-infrared part of the spectrum \citep{Hauser2001, Bernstein2002a, Bernstein2002b, Bernstein2002c, Matsumoto2005, Bernstein2007, Dwek2013}. This value is quite low and reflects the fact that sky is dark in the night.

Current theories assume that the low intensity of the EBL is caused primarily by finite age of the Universe and by its expansion \citep{Bondi1961, Wesson1987, Wesson1989, Knutsen1997}. It is argued according to the Olbers' paradox that the infinite static universe predicts a bright sky with intensity of the EBL comparable with surface brightness of stars being thus higher by more than 10 orders than the observed value. Consequently, the dark sky is taken as an important evidence for expanding universe of finite age. The finite age of the Universe implies that galaxies have not had time to populate the intergalactic space with enough photons to make it bright \citep{Wesson1989}. This argument is not, however, fully correct because it neglects an impact of light absorption by interstellar and intergalactic dust on the intensity of the EBL. 

The light extinction due to presence of absorbing interstellar dust has been observed, measured and numerically modelled by many authors \citep{Mathis1990, Charlot2000, Draine2003, Draine2011, Tuffs2004, Draine2007, Cunha2008, Somerville2012, Popescu2011}. The rate of light extinction is roughly 0.7-1.0 mag/kpc in the Milky Way \citep{Milne1980, Koppen1998}, but this value can vary being traced, for example, by dust temperature mapping \citep{Bernard2010}.  Obviously, light extinction is observed also in other galaxies and its rate depends on the type of the galaxy, its dust content and the galaxy inclination \citep{Goudfrooij1994, Calzetti2001, Holwerda2005a, Holwerda2007, Lisenfeld2008, Finkelman2008, Finkelman2010}. The light extinction due to interstellar dust affects the EBL in two ways. First, it reduces light radiated by galaxies and subsequently their surface brightness. Second, presence of dust in nearby foreground galaxies causes obscuration of distant background galaxies \citep{Gonzalez1998, Alton2001, Holwerda2005a, Holwerda2005b}. Consequently, the distant obscured galaxies become faint and do not contribute to the EBL significantly. In addition, the brightness of all distant galaxies is decreased because of light absorption by intergalactic dust. Obviously, accurate calculations of the EBL should take into account these effects. 

In this paper, we calculate the impact of light absorption by the galactic and intergalactic dust on the intensity of the stellar EBL. We show that the key factor responsible for the observed low stellar EBL is not the finite age or the expansion of the Universe \citep{Harrison1984, Harrison1990, Wesson1989, Knutsen1997} but a low surface brightness of galaxies and a partial opacity of galaxies and of the Universe due to absorbing dust.

\section{Light extinction}

Energy emitted by light sources and received at the Earth's surface is controlled by three basic factors: distance of light sources, extinction of light along a light ray by scattering and dust absorption, and obscuration of distant light sources by those at closer distance. The obscuration depends on the number of light sources in unit volume, their size and transparency. For fully opaque sources like stars, the obscuration is most effective, for partially opaque sources like galaxies, the obscuration is suppressed. The energy received at the Earth's surface is calculated by summing contributions of all light sources in a specified universe model. 

Energy flux $I$ received per unit area and time from light sources in the static universe is expressed as follows
\begin{equation}\label{eq1}
I = \iiint_V \frac{n L}{4\pi r^2} e^{-\left(\kappa/\gamma\right)r} e^{-\lambda r} dV
\end{equation}
where $r$ is the distance, $n$ is the mean number density of light sources (i.e., the mean number of light sources per unit volume), $L$ is the mean energy radiated by a light source per time (in W), $\kappa$ is the mean opacity of a light source, $\lambda$ is the mean absorption coefficient along a ray path, $dV$ is the volume element 
\begin{equation}\label{eq2}
dV = 4\pi r^2 dr
\end{equation}
and coefficient $\gamma$ is the mean free path of a light ray between light sources (i.e., the mean travelling distance of a photon emitted by one light source and reaching another light source)
\begin{equation}\label{eq3}
\gamma=\frac{1}{n \pi a^2}
\end{equation}
where $a$ is the mean radius of light sources. 	The opacity $\kappa$ in Eq. (1) quantifies how much energy is absorbed by a light source when external light goes through the source. Hence, $\kappa$ is 1 for a fully opaque galaxy and 0 for a fully transparent galaxy. Terms $e^{-\lambda r}$ and $e^{-\left(\kappa/\gamma\right)\,r}$ in Eq. (1) describe the light extinction and the obscuration \citep{Harrison1990, Knutsen1997} weighted by the opacity.

Integrating Eq. (1) we get
\begin{equation}\label{eq4}
I = n L \int_{r=0}^{\infty} e^{-\left(\kappa/\gamma + \lambda\right)\,r} dr = 
\frac{\gamma}{\kappa+\lambda \gamma} n L =\varepsilon j
\end{equation}
where
\begin{equation}\label{eq5}
\varepsilon=\frac{\gamma}{\kappa+\lambda \gamma}
\end{equation}
and
\begin{equation}\label{eq6}
j = nL
\end{equation}
is the luminosity density (in $\mathrm{W m}^{-3}$). Alternatively, Eq. (4) can be expressed as
\begin{equation}\label{eq7}
I = \frac{4l}{\kappa+\lambda \gamma}
\end{equation}
where
\begin{equation}\label{eq8}
l = \frac{L}{4\pi a^2}
\end{equation}
is the mean surface energy density (in $\mathrm{W m}^{-2}$) radiated by a light source. 

For a universe with uniformly distributed stars (opacity $\kappa$ of stars is 1) and for zero interstellar absorption, $\lambda = 0$, Eq. (7) yields the equality between the energy received at the unit area of the Earth $l_E$ and the mean surface energy $l_S$ radiated by a star (in $\mathrm{W m}^{-2}$) 
\begin{equation}\label{eq9}
l_E = l_S
\end{equation}
which is the mathematical formulation of the well-known Olbers' paradox \citep{Harrison1990, Knutsen1997}. 

For a universe with uniformly distributed galaxies with opacity $\kappa$ and for zero intergalactic absorption, $\lambda = 0$, Eq. (7) yields
\begin{equation}\label{eq10}
l_E = \frac{l_G}{\kappa}
\end{equation}
where $l_G$ is the mean surface energy radiated by a galaxy. Note that factor 4 in Eq. (7) is missing in Eqs. (9) and (10) because $l_E$ means the flux coming from the upper hemisphere and received at a unit area with a fixed (vertical) normal. Hence the integration in Eq. (1) is slightly different than in Eqs. (9) and (10) and yields a value which is four times lower \citep{Knutsen1997}.

In the case of fully transparent galaxies ($\kappa = 0$) in the fully transparent static universe ($\lambda = 0$), the received energy in Eq. (10) diverges. Obviously, this is not realistic, because stars, interstellar dust and intergalactic dust are fully or partially opaque, so $\kappa$ and $\lambda$ are apparently non-zero and cause the intensity of light observed at the Earth being low. For example, if 1\% of light energy is absorbed when the ray passes through a partially opaque galaxy and no energy is lost in the intergalactic space, Eq. (10) predicts the observed intensity of the EBL to be 100 times higher than the mean surface energy radiated by galaxies. Since the mean surface brightness of galaxies is extremely low, such intensity of the EBL corresponds effectively to the dark sky in the night.  

In order to calculate an accurate value of the intensity of the stellar EBL using eqs (4-6) we need values of number density $n$, radius of galaxies $a$, luminosity density $j$, galactic opacity $\kappa$ and intergalactic absorption $\lambda$ (also called the intergalactic or universe opacity). We shortly review estimates of these parameters based on observations in the next sections.

\begin{table}
%
\caption{Effective opacity of galaxies}  
\label{table:1}      
\centering                          
\begin{tabular}{c c c c}        
%
%
\hline\hline                 
%
%
Galaxy type & $A_V$ & $\kappa$ & $w$ \\  
%
&  (mag) &  & (\%) \\    
%
%
\hline                        
%
%
Elliptical & 0.06 $\pm$ 0.02 & 0.05  $\pm$ 0.02 & 35\\ 
Spiral     & 0.70 $\pm$ 0.20 & 0.48  $\pm$ 0.15 & 20\\
Lenticular & 0.30 $\pm$ 0.10 & 0.24  $\pm$ 0.08 & 45\\      
%
%
\hline  
\end{tabular}
%
%
%
\begin{tablenotes}
\item
$A_V$ is the effective inclination-averaged visual extinction,
$\kappa$ is the mean visual galactic opacity of the individual galaxy types, and 
$w$ is the relative frequency of the galaxy types taken from Table 4 (Typical galactic content of regular clusters) of \cite{Bahcall1999}.
\end{tablenotes}
\end{table}
%
\section{Number density, galaxy size and luminosity density}

The number density is fairly variable because of galaxy clustering and presence of voids in the universe \citep{Peebles2001, Jones2004, vonBenda_Beckmann2008}. The number density might be ten times higher or more in clusters, at distances up to 15-20 Mpc than the density averaged over larger distances. The mean value of the number density over hundreds of Mpc is, however, stable. It is derived from the Schechter luminosity function \citep{Schechter1976} being in the range of $0.010 - 0.025 \, h^3 \mathrm{Mpc}^{-3}$ \citep{Peebles1993, Peacock1999, Blanton2001, Blanton2003}. 

The size distribution of galaxies is dependent on their luminosity, stellar mass and the morphological type. Observed galaxies cover luminosities between $\sim10^8 L_{\sun}$ and $\sim10^{12} L_{\sun}$ with effective radii between $\sim 0.1\, h^{-1}$kpc and $\sim 10 \,h^{-1}$ kpc. For late-type galaxies, the characteristic luminosity in the R-band is -20.5 \citep{Shen2003}. The corresponding Petrosian half-light radius is $\sim 2.5-3.0\, h^{-1}$kpc and the $R90$ radius is about 3 times larger, $R90 = 7.5 - 9\, h^{-1}$kpc \citep{Graham2005}. This is close to a commonly assumed value $R = 10 \,h^{-1}$kpc \citep{Peebles1993, Peacock1999}.

The luminosity density is a fundamental quantity in observational cosmology standing in the Schechter luminosity function \citep{Schechter1976}. The most recent determination of the optical luminosity function come from large flux-limited redshift surveys such as the Two Degree Field Galaxy Redshift Survey (2dFGRS; \cite{Cross2001}), the Sloan Digital Sky Survey (SDSS; \cite{Blanton2001, Blanton2003}) or Century Survey (CS; \cite{Geller1997, Brown2001}). Independent estimates of the luminosity function in the R-band are well consistent being $(1.84 \pm 0.04) \times 10^8\, h\, L_{\sun}\, \mathrm{Mpc}^{-3}$ for the SDSS data \citep{Blanton2003} and $(1.9 \pm 0.6) \times 10^8 \,h L_{\sun}\, \mathrm{Mpc}^{-3}$ for the CS data \citep{Brown2001}.

%
\begin{table*}
%
\caption{Parameters for modelling of the stellar EBL}  
\label{Table:2}      
\centering                          
\begin{tabular}{c c c c c c c c}      
%
%
\hline\hline                 
%
%
 &  $n$ & $\gamma$ & $\kappa$ & $\lambda$ & $j^R$ & $I_{\mathrm{theor}}$ & $I_{\mathrm{obs}}$ \\  
%
   & (1/Mpc$^{3}$) & (Gpc) &  & (mag/Gpc) &  ($10^8\,L_{\sun} \,/\mathrm{Mpc}^{3}$) & (nW/m$^{2}$/sr) & (nW/m$^{2}$/sr) \\    
%
%
\hline                        
%
%
Minimum EBL & 0.015 & 130 & 0.30 & 0.03 & 1.80 & 190 &  20 \\      
Maximum EBL & 0.025 & 210 & 0.14 & 0.01 & 1.88 & 560 & 140 \\      
Optimum EBL & 0.020 & 160 & 0.22 & 0.02 & 1.84 & 290 &  60 \\      
%
%
\hline                                  
\end{tabular}
%
%
\begin{tablenotes}
\item
$n$ is the number density of galaxies,
$\gamma$ is the mean free path of light between galaxies defined in Eq. (3),
$\kappa$ is the mean opacity of galaxies,
$\lambda$ is the intergalactic absorption, and
$j^R$ is the R-band luminosity density \citep{Blanton2003},
$I_{\mathrm{theor}}$ is the predicted intensity of the stellar EBL, and
$I_{\mathrm{obs}}$ is the observed intensity of the stellar EBL.
The mean effective radius of galaxies $a$ is considered to be 10 kpc.
\end{tablenotes}
\end{table*}

\section{Galactic opacity}

The galactic opacity can be measured by a variety of methods usually applied to a large set of samples. The most widely used methods test dependence of the surface brightness on inclination, multi-wavelength comparisons, and statistical analysis of the colour and number count variations induced by a foreground galaxy onto background sources \citep{Calzetti2001}. 

The most transparent galaxies are elliptical. \cite{Goudfrooij1994} and \cite{Goudfrooij1995} found that the observed infrared luminosities are compatible with central optical depths of the diffuse component $\tau \left( 0 \right) \leq 0.7$, with a typical value of $\tau \left( 0 \right) \sim 0.2$. The corresponding effective extinction $A_V$ is $0.04 - 0.08$ mag. The giant elliptical galaxies found at the centres of cooling flow clusters are often surrounded by extended ($\approx 10 - 100$ kpc) and dusty emission-line nebulae \citep{Voit1997, Donahue2000}. Optical and UV emission-line studies give extinction values in the range $A_V \approx 0.3 - 2.0$ mag for the dust associated with the nebula. However, such galaxies are not statistically significant in the population of elliptical galaxies \citep{Calzetti2001}.

The dust extinction in spiral and irregular galaxies is higher than in elliptical galaxies. For estimating the extinction by dust in spiral galaxies, \cite{Holwerda2005b} used the so-called Synthetic field method (SFM) which counts the number of background galaxies seen through a foreground galaxy \citep{Gonzalez1998, Holwerda2005b}. The advantage of the SFM is that it yields the average opacity for the area of a galaxy disk without making assumptions about either the distribution of absorbers or of the disk starlight. \cite{Holwerda2005a} found that the dust opacity of the disk in the face-on view apparently arises from two distinct components: an optically thicker component ($A_I = 0.5 - 4$ mag) associated with the spiral arms and a relatively constant optically thinner disk ($A_I = 0.5$ mag). The early-type spiral disks show less extinction than the later types. As regards the inclination-averaged extinction, the typical values are according to \cite{Calzetti2001}: $0.5  - 0.75$ mag for Sa-Sab galaxies, $0.65  - 0.95$ mag for the Sb-Scd galaxies and $0.3  - 0.4$ mag for the irregular galaxies at the $B$-band.

In summary, galaxies in the local universe are moderately opaque, and extreme values of the opacity are found only in the statistically insignificant more active systems \citep{Calzetti2001}. Adopting estimates of the relative frequency of specific galaxy types in the Universe and their average extinctions  (see Table 1), we can calculate their mean visual opacities 
\begin{equation}\label{eq11}
\kappa = 1 - \exp \left(0.9211 A_V \right) \, ,
\end{equation}
and finally the overall mean galactic opacity using the weighted average  
\begin{equation}\label{eq12}
\kappa = w_1 \kappa_1 + w_2 \kappa_2 + w_3 \kappa_3 \, ,
\end{equation}
where subscripts 1, 2, and 3 stay for quantities of the elliptical, spiral and lenticular galaxies. According to Eq. (12) and Table 1, the average value of visual opacity $\kappa$ is about $0.22 \pm 0.08$. A more accurate approach should take into account the statistical distributions of the galaxy size and of the mean galaxy surface brightness for individual types of galaxies.

\section{Intergalactic opacity}

Observations indicate that the absorption of light is not limited to the interstellar medium within galaxies but is present also in the intergalactic space \citep{Nickerson1971, Margolis1977, Chelouche2007}. It is lower by several orders than in galaxies and depends on distance from galaxies. High values of attenuation are detected in galaxy halos \citep{Menard2010a} and in cluster centres. The attenuation due to dust in the galaxy clusters has been measured by reddening of background objects behind the clusters \citep{Chelouche2007, Bovy2008, Muller2008, Menard2010a}. The attenuation can also be investigated by correlations between the positions of high-redshift QSOs and low-redshift galaxies using catalogues of UVX objects. The excess of high-redshift QSOs around low-redshift galaxies is explained by a model in which dust situated in foreground clusters of galaxies obscures the QSOs lying behind them \citep{Boyle1988, Romani1992}.

Based on constructing extinction curves around galaxies, \cite{Menard2010a} found visual attenuation of $A_V = (1.3 \pm 0.1) \times 10^{-2}$ mag at distance from a galaxy up to 170  kpc and $A_V = (1.3 \pm 0.3) \times 10^{-3}$ mag on large scale at distance up to 1.7 Mpc. Values of the same order are reported for an average visual extinction by intracluster dust also by \cite{Muller2008} and 
\cite{Chelouche2007}. In addition, consistent opacity was recently reported by \cite{Xie2015} who studied the luminosity and redshifts of the quasar continuum at the data sample of $\sim 90.000$ objects and estimated the effective dust density $n\sigma_V \approx 0.02\, h\, \mathrm{Gpc}^{-1}$ at $z < 1.5$. However, the intergalactic absorption is redshift dependent. According to \cite{Davies1997} the intergalactic extinction increases with redshift and transparent universe becomes significantly opaque (optically thick) at redshifts of $z = 1 - 3$. The increase of intergalactic extinction with redshift is confirmed by results of \cite{Menard2010a} who estimated $A_V$ to about 0.03 mag at $z = 0.5$ but to about $0.05 - 0.09$ mag at $z = 1$.

\section{Predicted and observed stellar EBL}
\begin{figure*}
\centering
\includegraphics[angle=0,width=15cm,trim = {125 180 140 135},clip]{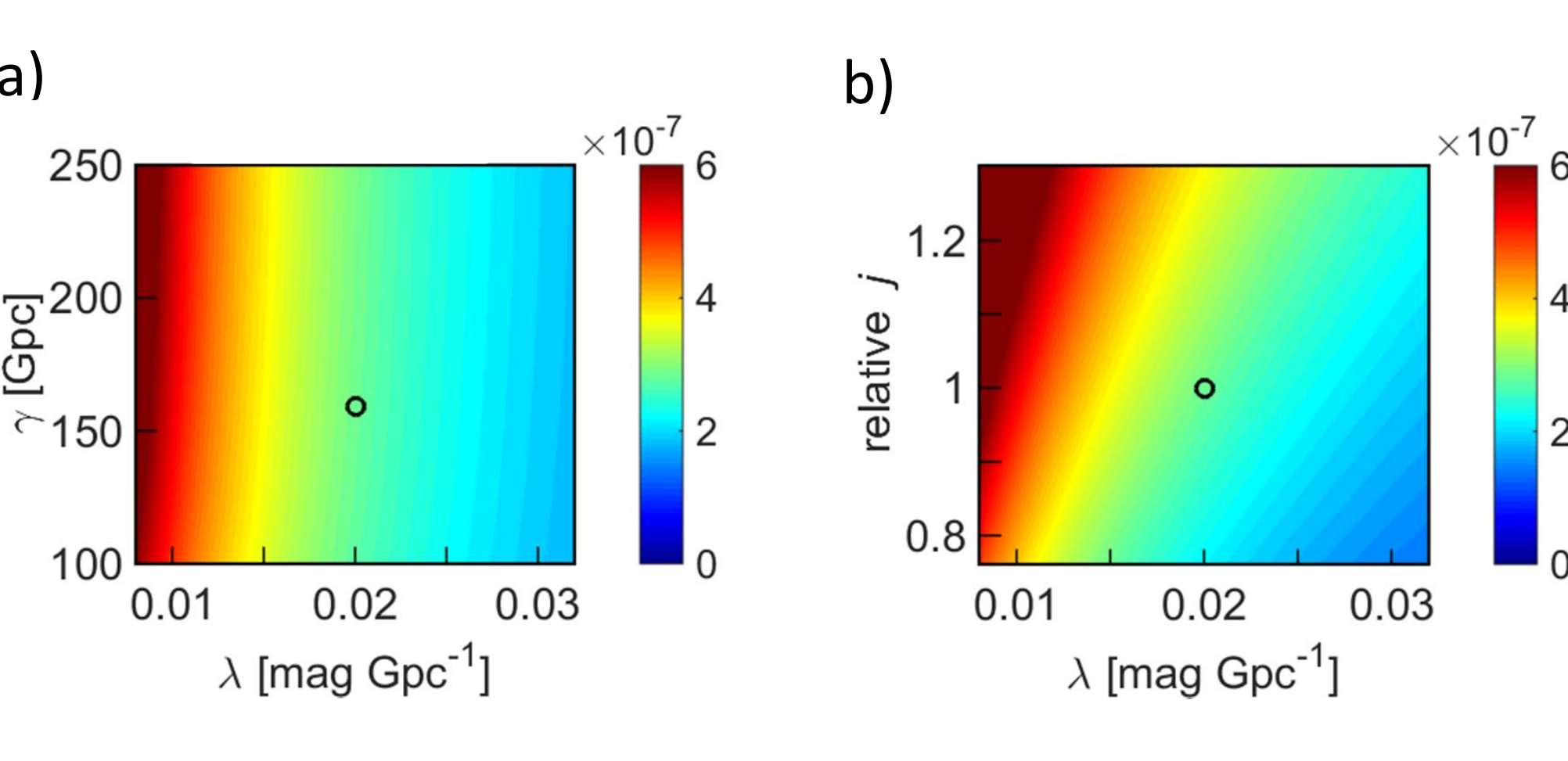}
\caption{
(a) The stellar EBL as a function of intergalactic opacity $\lambda$ and mean free path between galaxies $\gamma$. (b) The stellar EBL as a function of the intergalactic opacity $\lambda$ and  relative luminosity density $j$. The relative luminosity density is normalized to its optimum value (see Tab. 2). The black open circles mark the positions of optimally chosen parameters. The colour-coded EBL is in $\mathrm{W m}^{-2} \mathrm{sr}^{-1}$. 
}
\label{fig:2}
\end{figure*}

Taking into account estimates of the galactic and intergalactic opacity and other cosmological parameters (see Table 2), the intensity of the stellar EBL calculated by Eqs. (4) and (5) lies for wavelengths between 300 and 3500 nm in the range of $190-560 \,\mathrm{n W m}^{-2}\mathrm{sr}^{-1}$ with the optimum value of
\begin{equation}\label{eq13}
I_{\mathrm{theor}} \approx 290 \,\mathrm{n W m}^{-2}\mathrm{sr}^{-1} \, .
\end{equation}
Fig. 2 indicates that the EBL is rather insensitive to mean free path between galaxies $\gamma$ but quite sensitive to intergalactic opacity $\lambda$ and luminosity density $j$. Obviously, high values of the EBL are produced for high luminosity density and low intergalactic opacity. 

The observed intensity of the stellar EBL \citep{Hauser2001, Bernstein2002a, Bernstein2002b, Bernstein2002c, Bernstein2007} lies in the range of $20-140 \,\mathrm{n W m}^{-2}\mathrm{sr}^{-1}$ (see Fig. 1) with the optimum value of
\begin{equation}\label{eq14}
I_{\mathrm{obs}} \approx 60 \,\mathrm{n W m}^{-2}\mathrm{sr}^{-1} \, .
\end{equation}
Hence the predicted stellar EBL is about 5 times higher than the observed EBL. This result is surprising because it is commonly assumed that the EBL calculated for the infinite static universe must be higher than the observed EBL by more than 10 orders. The rather low value of the EBL in Eq. (13) evidences that the key factor for a successful prediction of the EBL is including the effects of the galactic and intergalactic absorption of light by dust. In fact, considering an expanding universe in the EBL calculations is not essential. Substituting the infinite static universe by expanding universe of finite age in the EBL calculations further reduces the predicted EBL but by factor of 5 only.

\section{Obscuration by galaxies versus intergalactic opacity}

\begin{figure}
\centering
\includegraphics[angle=0,width = 8.0 cm, trim = {310 40 310 40},clip]{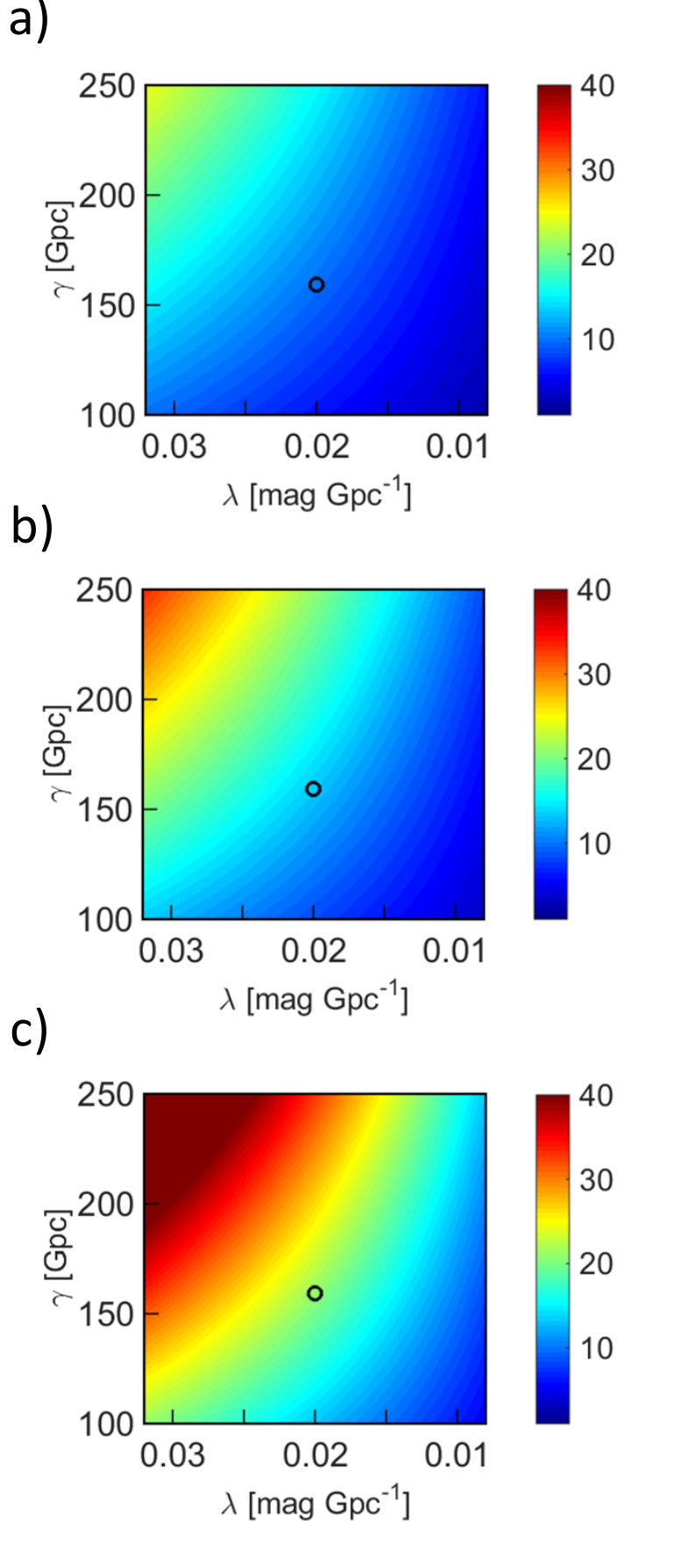}
%
\caption{
Ratio $k$ evaluating the impact of the intergalactic opacity and the obscuration of galaxies on the stellar EBL. The ratio is shown as a function of intergalactic opacity $\lambda$ and mean free path between galaxies $\gamma$. The galactic opacity is: (a) $\kappa = 0.30$, (b) $\kappa = 0.22$, and (c) $\kappa = 0.14$. Considering optimum values of $\lambda$ and $\gamma$, ratio $k$ is: (a) $k = 10$, (b) $k = 13$, and (c) $k = 21$.
}
\label{fig:3}
\end{figure}
 
As shown above, the low value of the stellar EBL is caused by two effects: (1) obscuration of background galaxies by partially opaque foreground galaxies, and (2) the intergalactic opacity. Considering observations of the number density of galaxies and of the galactic and intergalactic opacity, we can determine which effect has a more significant impact on the EBL. We calculate ratio $k$
\begin{equation}\label{eq15}
k = \frac{\lambda \gamma}{\kappa} \, ,
\end{equation}
that is higher (lower) than 1 for the intergalactic opacity (obscuration of galaxies) being predominant.

Figure 3 shows ratio $k$ as a function of intergalactic opacity $\lambda$ and mean free path $\gamma$. The ratio is calculated for three values of the mean galactic opacity, $\kappa = 0.30$, 0.22 and 0.14. The corresponding values of $k$ for optimum values of $\lambda$ and $\kappa$ are 10, 13 and 21. This evidences that the EBL is affected predominantly by the intergalactic opacity. The impact of the obscuration effect on the EBL is almost negligible for the majority of combinations of realistic values of $\lambda$ and $\kappa$.

\section{Discussion}

The calculations prove that the impact of light absorption by galactic and intergalactic dust on the stellar EBL is significant. It is even more significant than the expansion, redshift or the finite age of the Universe. The absorbed starlight heats up the dust and is further re-radiated at the IR, FIR and micro-wave spectrum. Although the presented calculations are rough approximations, the estimate of the stellar EBL is robust and reliable. It is based on: (1) the luminosity density measurements \citep{Blanton2001, Blanton2003, Brown2001}, (2) the estimate of the effective inclination-averaged galactic opacity \citep{Calzetti2001}, and (3) the estimate of the intergalactic opacity \citep{Xie2015}. Even a minute intergalactic opacity of $1 \times 10^{-2} \, \mathrm{mag \, Gpc}^{-1}$ is high enough to produce significant effects on the EBL. In addition, galaxies are partially opaque due to galactic dust. Consequently, the opacity of foreground galaxies reduces the intensity of light radiated by distant background galaxies. The background galaxies become apparently faint and do not contribute to the stellar EBL significantly. As a result, the stellar EBL is comparable with or even lower than the mean surface brightness of galaxies. Comparing the impact of the intergalactic opacity and obscuration of galaxies on the EBL, the intergalactic opacity is more significant by factor of 10 or more. If the intergalactic opacity and the obscuration effects are neglected, the EBL predicted for the static universe should be according to the Olbers’ paradox as high as the surface brightness of stars. This value is higher by almost 14 orders than the observed one. Hence, the galactic and intergalactic absorption is the most important factor for the observed low stellar EBL. The corrections for the expansion of the Universe and for the redshift alter the predictions by less than one order only. 


\acknowledgments
I thank Benne W. Holwerda for his detailed and very helpful comments on the paper and Alberto Dom\'{i}nguez for providing me kindly with Fig. 1.


\bibliographystyle{spr-mp-nameyear-cnd}  
\bibliography{paper1}                     

\begin{thebibliography}{72}
\ifx \bisbn   \undefined \def \bisbn  #1{ISBN #1}\fi
\ifx \binits  \undefined \def \binits#1{#1} \fi
\ifx \bauthor  \undefined \def \bauthor#1{#1} \fi
\ifx \batitle  \undefined \def \batitle#1{#1} \fi
\ifx \bjtitle  \undefined \def \bjtitle#1{#1}\fi
\ifx \bvolume  \undefined \def \bvolume#1{\textbf{#1}}\fi
\ifx \byear  \undefined \def \byear#1{#1} \fi
\ifx \bissue  \undefined \def \bissue#1{#1} \fi
\ifx \bfpage  \undefined \def \bfpage#1{#1} \fi
\ifx \blpage  \undefined \def \blpage #1{#1} \fi
\ifx \burl  \undefined \def \burl#1{\textsf{#1}} \fi
\ifx \doiurl  \undefined \def \doiurl#1{\textsf{#1}} \fi
\ifx \betal  \undefined \def \betal{\textit{et al.}} \fi
\ifx \binstitute  \undefined \def \binstitute#1{#1} \fi
\ifx \binstitutionaled  \undefined \def \binstitutionaled#1{#1} \fi
\ifx \bctitle  \undefined \def \bctitle#1{#1} \fi
\ifx \beditor  \undefined \def \beditor#1{#1} \fi
\ifx \bpublisher  \undefined \def \bpublisher#1{#1} \fi
\ifx \bbtitle  \undefined \def \bbtitle#1{#1} \fi
\ifx \bedition  \undefined \def \bedition#1{#1} \fi
\ifx \bseriesno  \undefined \def \bseriesno#1{#1} \fi
\ifx \blocation  \undefined \def \blocation#1{#1} \fi
\ifx \bsertitle  \undefined \def \bsertitle#1{#1} \fi
\ifx \bsnm \undefined \def \bsnm#1{#1} \fi
\ifx \bsuffix \undefined \def \bsuffix#1{#1} \fi
\ifx \bparticle \undefined \def \bparticle#1{#1} \fi
\ifx \barticle \undefined \def \barticle#1{#1} \fi
\ifx \bconfdate \undefined \def \bconfdate #1{#1} \fi
\ifx \botherref \undefined \def \botherref #1{#1} \fi
\ifx \url \undefined \def \url#1{\textsf{#1}} \fi
\ifx \bchapter \undefined \def \bchapter#1{#1} \fi
\ifx \bbook \undefined \def \bbook#1{#1} \fi
\ifx \bcomment \undefined \def \bcomment#1{#1} \fi
\ifx \oauthor \undefined \def \oauthor#1{#1} \fi
\ifx \citeauthoryear \undefined \def \citeauthoryear#1{#1} \fi
\ifx \endbibitem  \undefined \def \endbibitem {}\fi
\ifx \bconflocation  \undefined \def \bconflocation#1{#1} \fi
\ifx \arxivurl  \undefined \def \arxivurl#1{\textsf{#1}} \fi

\bibitem[\protect\citeauthoryear{{Alton} et~al.}{2001}]{Alton2001}
\begin{barticle}
\bauthor{\bsnm{{Alton}}, \binits{P.B.}},
\bauthor{\bsnm{{Bianchi}}, \binits{S.}},
\bauthor{\bsnm{{Davies}}, \binits{J.}}:
\bjtitle{\apss}
\bvolume{276},
\bfpage{949}
(\byear{2001}).
doi:\doiurl{10.1023/A:1017535123314}
\end{barticle}
\endbibitem

\bibitem[\protect\citeauthoryear{{Bahcall}}{1999}]{Bahcall1999}
\begin{bchapter}
\bauthor{\bsnm{{Bahcall}}, \binits{N.A.}}:
In: \beditor{\bsnm{{Dekel}}, \binits{A.}},
\beditor{\bsnm{{Ostriker}}, \binits{J.P.}} (eds.)
\bbtitle{Formation of Structure in the Universe},
p. \bfpage{135}
(\byear{1999})
\end{bchapter}
\endbibitem

\bibitem[\protect\citeauthoryear{{Bernard} et~al.}{2010}]{Bernard2010}
\begin{barticle}
\bauthor{\bsnm{{Bernard}}, \binits{J.-P.}},
\bauthor{\bsnm{{Paradis}}, \binits{D.}},
\bauthor{\bsnm{{Marshall}}, \binits{D.J.}},
\bauthor{\bsnm{{Montier}}, \binits{L.}},
\bauthor{\bsnm{{Lagache}}, \binits{G.}},
\bauthor{\bsnm{{Paladini}}, \binits{R.}},
\bauthor{\bsnm{{Veneziani}}, \binits{M.}},
\bauthor{\bsnm{{Brunt}}, \binits{C.M.}},
\bauthor{\bsnm{{Mottram}}, \binits{J.C.}},
\bauthor{\bsnm{{Martin}}, \binits{P.}},
\bauthor{\bsnm{{Ristorcelli}}, \binits{I.}},
\bauthor{\bsnm{{Noriega-Crespo}}, \binits{A.}},
\bauthor{\bsnm{{Compi{\`e}gne}}, \binits{M.}},
\bauthor{\bsnm{{Flagey}}, \binits{N.}},
\bauthor{\bsnm{{Anderson}}, \binits{L.D.}},
\bauthor{\bsnm{{Popescu}}, \binits{C.C.}},
\bauthor{\bsnm{{Tuffs}}, \binits{R.}},
\bauthor{\bsnm{{Reach}}, \binits{W.}},
\bauthor{\bsnm{{White}}, \binits{G.}},
\bauthor{\bsnm{{Benedettini}}, \binits{M.}},
\bauthor{\bsnm{{Calzoletti}}, \binits{L.}},
\bauthor{\bsnm{{Digiorgio}}, \binits{A.M.}},
\bauthor{\bsnm{{Faustini}}, \binits{F.}},
\bauthor{\bsnm{{Juvela}}, \binits{M.}},
\bauthor{\bsnm{{Joblin}}, \binits{C.}},
\bauthor{\bsnm{{Joncas}}, \binits{G.}},
\bauthor{\bsnm{{Mivilles-Deschenes}}, \binits{M.-A.}},
\bauthor{\bsnm{{Olmi}}, \binits{L.}},
\bauthor{\bsnm{{Traficante}}, \binits{A.}},
\bauthor{\bsnm{{Piacentini}}, \binits{F.}},
\bauthor{\bsnm{{Zavagno}}, \binits{A.}},
\bauthor{\bsnm{{Molinari}}, \binits{S.}}:
\bjtitle{\aap}
\bvolume{518},
\bfpage{88}
(\byear{2010}).
doi:\doiurl{10.1051/0004-6361/201014540}
\end{barticle}
\endbibitem

\bibitem[\protect\citeauthoryear{{Bernstein}}{2007}]{Bernstein2007}
\begin{barticle}
\bauthor{\bsnm{{Bernstein}}, \binits{R.A.}}:
\bjtitle{\apj}
\bvolume{666},
\bfpage{663}
(\byear{2007})
\end{barticle}
\endbibitem

\bibitem[\protect\citeauthoryear{{Bernstein} et~al.}{2002a}]{Bernstein2002a}
\begin{barticle}
\bauthor{\bsnm{{Bernstein}}, \binits{R.A.}},
\bauthor{\bsnm{{Freedman}}, \binits{W.L.}},
\bauthor{\bsnm{{Madore}}, \binits{B.F.}}:
\bjtitle{\apj}
\bvolume{571},
\bfpage{56}
(\byear{2002}a).
doi:\doiurl{10.1086/339422}
\end{barticle}
\endbibitem

\bibitem[\protect\citeauthoryear{{Bernstein} et~al.}{2002b}]{Bernstein2002b}
\begin{barticle}
\bauthor{\bsnm{{Bernstein}}, \binits{R.A.}},
\bauthor{\bsnm{{Freedman}}, \binits{W.L.}},
\bauthor{\bsnm{{Madore}}, \binits{B.F.}}:
\bjtitle{\apj}
\bvolume{571},
\bfpage{85}
(\byear{2002}b).
doi:\doiurl{10.1086/339423}
\end{barticle}
\endbibitem

\bibitem[\protect\citeauthoryear{{Bernstein} et~al.}{2002c}]{Bernstein2002c}
\begin{barticle}
\bauthor{\bsnm{{Bernstein}}, \binits{R.A.}},
\bauthor{\bsnm{{Freedman}}, \binits{W.L.}},
\bauthor{\bsnm{{Madore}}, \binits{B.F.}}:
\bjtitle{\apj}
\bvolume{571},
\bfpage{107}
(\byear{2002}c).
doi:\doiurl{10.1086/339424}
\end{barticle}
\endbibitem

\bibitem[\protect\citeauthoryear{{Biteau} and {Williams}}{2015}]{Biteau2015}
\begin{barticle}
\bauthor{\bsnm{{Biteau}}, \binits{J.}},
\bauthor{\bsnm{{Williams}}, \binits{D.A.}}:
\bjtitle{\apj}
\bvolume{812},
\bfpage{60}
(\byear{2015}).
doi:\doiurl{10.1088/0004-637X/812/1/60}
\end{barticle}
\endbibitem

\bibitem[\protect\citeauthoryear{{Blanton} et~al.}{2001}]{Blanton2001}
\begin{barticle}
\bauthor{\bsnm{{Blanton}}, \binits{M.R.}},
\bauthor{\bsnm{{Dalcanton}}, \binits{J.}},
\bauthor{\bsnm{{Eisenstein}}, \binits{D.}},
\bauthor{\bsnm{{Loveday}}, \binits{J.}},
\bauthor{\bsnm{{Strauss}}, \binits{M.A.}},
\bauthor{\bsnm{{SubbaRao}}, \binits{M.}},
\bauthor{\bsnm{{Weinberg}}, \binits{D.H.}},
\bauthor{\bsnm{{Anderson}}, \binits{J.E.} \bsuffix{Jr.}},
\bauthor{\bsnm{{Annis}}, \binits{J.}},
\bauthor{\bsnm{{Bahcall}}, \binits{N.A.}},
\bauthor{\bsnm{{Bernardi}}, \binits{M.}},
\bauthor{\bsnm{{Brinkmann}}, \binits{J.}},
\bauthor{\bsnm{{Brunner}}, \binits{R.J.}},
\bauthor{\bsnm{{Burles}}, \binits{S.}},
\bauthor{\bsnm{{Carey}}, \binits{L.}},
\bauthor{\bsnm{{Castander}}, \binits{F.J.}},
\bauthor{\bsnm{{Connolly}}, \binits{A.J.}},
\bauthor{\bsnm{{Csabai}}, \binits{I.}},
\bauthor{\bsnm{{Doi}}, \binits{M.}},
\bauthor{\bsnm{{Finkbeiner}}, \binits{D.}},
\bauthor{\bsnm{{Friedman}}, \binits{S.}},
\bauthor{\bsnm{{Frieman}}, \binits{J.A.}},
\bauthor{\bsnm{{Fukugita}}, \binits{M.}},
\bauthor{\bsnm{{Gunn}}, \binits{J.E.}},
\bauthor{\bsnm{{Hennessy}}, \binits{G.S.}},
\bauthor{\bsnm{{Hindsley}}, \binits{R.B.}},
\bauthor{\bsnm{{Hogg}}, \binits{D.W.}},
\bauthor{\bsnm{{Ichikawa}}, \binits{T.}},
\bauthor{\bsnm{{Ivezi{\'c}}}, \binits{{\v Z}.}},
\bauthor{\bsnm{{Kent}}, \binits{S.}},
\bauthor{\bsnm{{Knapp}}, \binits{G.R.}},
\bauthor{\bsnm{{Lamb}}, \binits{D.Q.}},
\bauthor{\bsnm{{Leger}}, \binits{R.F.}},
\bauthor{\bsnm{{Long}}, \binits{D.C.}},
\bauthor{\bsnm{{Lupton}}, \binits{R.H.}},
\bauthor{\bsnm{{McKay}}, \binits{T.A.}},
\bauthor{\bsnm{{Meiksin}}, \binits{A.}},
\bauthor{\bsnm{{Merelli}}, \binits{A.}},
\bauthor{\bsnm{{Munn}}, \binits{J.A.}},
\bauthor{\bsnm{{Narayanan}}, \binits{V.}},
\bauthor{\bsnm{{Newcomb}}, \binits{M.}},
\bauthor{\bsnm{{Nichol}}, \binits{R.C.}},
\bauthor{\bsnm{{Okamura}}, \binits{S.}},
\bauthor{\bsnm{{Owen}}, \binits{R.}},
\bauthor{\bsnm{{Pier}}, \binits{J.R.}},
\bauthor{\bsnm{{Pope}}, \binits{A.}},
\bauthor{\bsnm{{Postman}}, \binits{M.}},
\bauthor{\bsnm{{Quinn}}, \binits{T.}},
\bauthor{\bsnm{{Rockosi}}, \binits{C.M.}},
\bauthor{\bsnm{{Schlegel}}, \binits{D.J.}},
\bauthor{\bsnm{{Schneider}}, \binits{D.P.}},
\bauthor{\bsnm{{Shimasaku}}, \binits{K.}},
\bauthor{\bsnm{{Siegmund}}, \binits{W.A.}},
\bauthor{\bsnm{{Smee}}, \binits{S.}},
\bauthor{\bsnm{{Snir}}, \binits{Y.}},
\bauthor{\bsnm{{Stoughton}}, \binits{C.}},
\bauthor{\bsnm{{Stubbs}}, \binits{C.}},
\bauthor{\bsnm{{Szalay}}, \binits{A.S.}},
\bauthor{\bsnm{{Szokoly}}, \binits{G.P.}},
\bauthor{\bsnm{{Thakar}}, \binits{A.R.}},
\bauthor{\bsnm{{Tremonti}}, \binits{C.}},
\bauthor{\bsnm{{Tucker}}, \binits{D.L.}},
\bauthor{\bsnm{{Uomoto}}, \binits{A.}},
\bauthor{\bsnm{{Vanden Berk}}, \binits{D.}},
\bauthor{\bsnm{{Vogeley}}, \binits{M.S.}},
\bauthor{\bsnm{{Waddell}}, \binits{P.}},
\bauthor{\bsnm{{Yanny}}, \binits{B.}},
\bauthor{\bsnm{{Yasuda}}, \binits{N.}},
\bauthor{\bsnm{{York}}, \binits{D.G.}}:
\bjtitle{\aj}
\bvolume{121},
\bfpage{2358}
(\byear{2001}).
\arxivurl{astro-ph/0012085}.
doi:\doiurl{10.1086/320405}
\end{barticle}
\endbibitem

\bibitem[\protect\citeauthoryear{{Blanton} et~al.}{2003}]{Blanton2003}
\begin{barticle}
\bauthor{\bsnm{{Blanton}}, \binits{M.R.}},
\bauthor{\bsnm{{Hogg}}, \binits{D.W.}},
\bauthor{\bsnm{{Bahcall}}, \binits{N.A.}},
\bauthor{\bsnm{{Brinkmann}}, \binits{J.}},
\bauthor{\bsnm{{Britton}}, \binits{M.}},
\bauthor{\bsnm{{Connolly}}, \binits{A.J.}},
\bauthor{\bsnm{{Csabai}}, \binits{I.}},
\bauthor{\bsnm{{Fukugita}}, \binits{M.}},
\bauthor{\bsnm{{Loveday}}, \binits{J.}},
\bauthor{\bsnm{{Meiksin}}, \binits{A.}},
\bauthor{\bsnm{{Munn}}, \binits{J.A.}},
\bauthor{\bsnm{{Nichol}}, \binits{R.C.}},
\bauthor{\bsnm{{Okamura}}, \binits{S.}},
\bauthor{\bsnm{{Quinn}}, \binits{T.}},
\bauthor{\bsnm{{Schneider}}, \binits{D.P.}},
\bauthor{\bsnm{{Shimasaku}}, \binits{K.}},
\bauthor{\bsnm{{Strauss}}, \binits{M.A.}},
\bauthor{\bsnm{{Tegmark}}, \binits{M.}},
\bauthor{\bsnm{{Vogeley}}, \binits{M.S.}},
\bauthor{\bsnm{{Weinberg}}, \binits{D.H.}}:
\bjtitle{\apj}
\bvolume{592},
\bfpage{819}
(\byear{2003}).
\arxivurl{astro-ph/0210215}.
doi:\doiurl{10.1086/375776}
\end{barticle}
\endbibitem

\bibitem[\protect\citeauthoryear{{Bondi}}{1961}]{Bondi1961}
\begin{bbook}
\bauthor{\bsnm{{Bondi}}, \binits{H.}}:
\bbtitle{{Cosmology}},
(\byear{1961})
\end{bbook}
\endbibitem

\bibitem[\protect\citeauthoryear{{Bovy} et~al.}{2008}]{Bovy2008}
\begin{barticle}
\bauthor{\bsnm{{Bovy}}, \binits{J.}},
\bauthor{\bsnm{{Hogg}}, \binits{D.W.}},
\bauthor{\bsnm{{Moustakas}}, \binits{J.}}:
\bjtitle{\apj}
\bvolume{688},
\bfpage{198}
(\byear{2008}).
\arxivurl{0805.1200}.
doi:\doiurl{10.1086/592187}
\end{barticle}
\endbibitem

\bibitem[\protect\citeauthoryear{{Boyle} et~al.}{1988}]{Boyle1988}
\begin{barticle}
\bauthor{\bsnm{{Boyle}}, \binits{B.J.}},
\bauthor{\bsnm{{Fong}}, \binits{R.}},
\bauthor{\bsnm{{Shanks}}, \binits{T.}}:
\bjtitle{\mnras}
\bvolume{231},
\bfpage{897}
(\byear{1988})
\end{barticle}
\endbibitem

\bibitem[\protect\citeauthoryear{{Brown} et~al.}{2001}]{Brown2001}
\begin{barticle}
\bauthor{\bsnm{{Brown}}, \binits{W.R.}},
\bauthor{\bsnm{{Geller}}, \binits{M.J.}},
\bauthor{\bsnm{{Fabricant}}, \binits{D.G.}},
\bauthor{\bsnm{{Kurtz}}, \binits{M.J.}}:
\bjtitle{\aj}
\bvolume{122},
\bfpage{714}
(\byear{2001}).
doi:\doiurl{10.1086/321176}
\end{barticle}
\endbibitem

\bibitem[\protect\citeauthoryear{{Calzetti}}{2001}]{Calzetti2001}
\begin{barticle}
\bauthor{\bsnm{{Calzetti}}, \binits{D.}}:
\bjtitle{\pasp}
\bvolume{113},
\bfpage{1449}
(\byear{2001}).
\arxivurl{astro-ph/0109035}.
doi:\doiurl{10.1086/324269}
\end{barticle}
\endbibitem

\bibitem[\protect\citeauthoryear{{Calzetti} et~al.}{2000}]{Calzetti2000}
\begin{barticle}
\bauthor{\bsnm{{Calzetti}}, \binits{D.}},
\bauthor{\bsnm{{Armus}}, \binits{L.}},
\bauthor{\bsnm{{Bohlin}}, \binits{R.C.}},
\bauthor{\bsnm{{Kinney}}, \binits{A.L.}},
\bauthor{\bsnm{{Koornneef}}, \binits{J.}},
\bauthor{\bsnm{{Storchi-Bergmann}}, \binits{T.}}:
\bjtitle{\apj}
\bvolume{533},
\bfpage{682}
(\byear{2000}).
\arxivurl{astro-ph/9911459}.
doi:\doiurl{10.1086/308692}
\end{barticle}
\endbibitem

\bibitem[\protect\citeauthoryear{{Charlot} and {Fall}}{2000}]{Charlot2000}
\begin{barticle}
\bauthor{\bsnm{{Charlot}}, \binits{S.}},
\bauthor{\bsnm{{Fall}}, \binits{S.M.}}:
\bjtitle{\apj}
\bvolume{539},
\bfpage{718}
(\byear{2000}).
\arxivurl{astro-ph/0003128}.
doi:\doiurl{10.1086/309250}
\end{barticle}
\endbibitem

\bibitem[\protect\citeauthoryear{{Chelouche} et~al.}{2007}]{Chelouche2007}
\begin{barticle}
\bauthor{\bsnm{{Chelouche}}, \binits{D.}},
\bauthor{\bsnm{{Koester}}, \binits{B.P.}},
\bauthor{\bsnm{{Bowen}}, \binits{D.V.}}:
\bjtitle{\apjl}
\bvolume{671},
\bfpage{97}
(\byear{2007}).
\arxivurl{0711.1167}.
doi:\doiurl{10.1086/525251}
\end{barticle}
\endbibitem

\bibitem[\protect\citeauthoryear{{Cross} et~al.}{2001}]{Cross2001}
\begin{barticle}
\bauthor{\bsnm{{Cross}}, \binits{N.}},
\bauthor{\bsnm{{Driver}}, \binits{S.P.}},
\bauthor{\bsnm{{Couch}}, \binits{W.}},
\bauthor{\bsnm{{Baugh}}, \binits{C.M.}},
\bauthor{\bsnm{{Bland-Hawthorn}}, \binits{J.}},
\bauthor{\bsnm{{Bridges}}, \binits{T.}},
\bauthor{\bsnm{{Cannon}}, \binits{R.}},
\bauthor{\bsnm{{Cole}}, \binits{S.}},
\bauthor{\bsnm{{Colless}}, \binits{M.}},
\bauthor{\bsnm{{Collins}}, \binits{C.}},
\bauthor{\bsnm{{Dalton}}, \binits{G.}},
\bauthor{\bsnm{{Deeley}}, \binits{K.}},
\bauthor{\bsnm{{De Propris}}, \binits{R.}},
\bauthor{\bsnm{{Efstathiou}}, \binits{G.}},
\bauthor{\bsnm{{Ellis}}, \binits{R.S.}},
\bauthor{\bsnm{{Frenk}}, \binits{C.S.}},
\bauthor{\bsnm{{Glazebrook}}, \binits{K.}},
\bauthor{\bsnm{{Jackson}}, \binits{C.}},
\bauthor{\bsnm{{Lahav}}, \binits{O.}},
\bauthor{\bsnm{{Lewis}}, \binits{I.}},
\bauthor{\bsnm{{Lumsden}}, \binits{S.}},
\bauthor{\bsnm{{Maddox}}, \binits{S.}},
\bauthor{\bsnm{{Madgwick}}, \binits{D.}},
\bauthor{\bsnm{{Moody}}, \binits{S.}},
\bauthor{\bsnm{{Norberg}}, \binits{P.}},
\bauthor{\bsnm{{Peacock}}, \binits{J.A.}},
\bauthor{\bsnm{{Peterson}}, \binits{B.A.}},
\bauthor{\bsnm{{Price}}, \binits{I.}},
\bauthor{\bsnm{{Seaborne}}, \binits{M.}},
\bauthor{\bsnm{{Sutherland}}, \binits{W.}},
\bauthor{\bsnm{{Tadros}}, \binits{H.}},
\bauthor{\bsnm{{Taylor}}, \binits{K.}}:
\bjtitle{\mnras}
\bvolume{324},
\bfpage{825}
(\byear{2001}).
\arxivurl{astro-ph/0012165}.
doi:\doiurl{10.1046/j.1365-8711.2001.04254.x}
\end{barticle}
\endbibitem

\bibitem[\protect\citeauthoryear{{da Cunha} et~al.}{2008}]{Cunha2008}
\begin{barticle}
\bauthor{\bsnm{{da Cunha}}, \binits{E.}},
\bauthor{\bsnm{{Charlot}}, \binits{S.}},
\bauthor{\bsnm{{Elbaz}}, \binits{D.}}:
\bjtitle{\mnras}
\bvolume{388},
\bfpage{1595}
(\byear{2008}).
\arxivurl{0806.1020}.
doi:\doiurl{10.1111/j.1365-2966.2008.13535.x}
\end{barticle}
\endbibitem

\bibitem[\protect\citeauthoryear{{Davies} et~al.}{1997}]{Davies1997}
\begin{barticle}
\bauthor{\bsnm{{Davies}}, \binits{J.I.}},
\bauthor{\bsnm{{Phillipps}}, \binits{S.}},
\bauthor{\bsnm{{Trewhella}}, \binits{M.}},
\bauthor{\bsnm{{Alton}}, \binits{P.}}:
\bjtitle{\mnras}
\bvolume{291},
\bfpage{59}
(\byear{1997})
\end{barticle}
\endbibitem

\bibitem[\protect\citeauthoryear{{Dom{\'{\i}}nguez}
  et~al.}{2011}]{Dominguez2011}
\begin{barticle}
\bauthor{\bsnm{{Dom{\'{\i}}nguez}}, \binits{A.}},
\bauthor{\bsnm{{Primack}}, \binits{J.R.}},
\bauthor{\bsnm{{Rosario}}, \binits{D.J.}},
\bauthor{\bsnm{{Prada}}, \binits{F.}},
\bauthor{\bsnm{{Gilmore}}, \binits{R.C.}},
\bauthor{\bsnm{{Faber}}, \binits{S.M.}},
\bauthor{\bsnm{{Koo}}, \binits{D.C.}},
\bauthor{\bsnm{{Somerville}}, \binits{R.S.}},
\bauthor{\bsnm{{P{\'e}rez-Torres}}, \binits{M.A.}},
\bauthor{\bsnm{{P{\'e}rez-Gonz{\'a}lez}}, \binits{P.}},
\bauthor{\bsnm{{Huang}}, \binits{J.-S.}},
\bauthor{\bsnm{{Davis}}, \binits{M.}},
\bauthor{\bsnm{{Guhathakurta}}, \binits{P.}},
\bauthor{\bsnm{{Barmby}}, \binits{P.}},
\bauthor{\bsnm{{Conselice}}, \binits{C.J.}},
\bauthor{\bsnm{{Lozano}}, \binits{M.}},
\bauthor{\bsnm{{Newman}}, \binits{J.A.}},
\bauthor{\bsnm{{Cooper}}, \binits{M.C.}}:
\bjtitle{\mnras}
\bvolume{410},
\bfpage{2556}
(\byear{2011}).
\arxivurl{1007.1459}.
doi:\doiurl{10.1111/j.1365-2966.2010.17631.x}
\end{barticle}
\endbibitem

\bibitem[\protect\citeauthoryear{{Donahue} et~al.}{2000}]{Donahue2000}
\begin{barticle}
\bauthor{\bsnm{{Donahue}}, \binits{M.}},
\bauthor{\bsnm{{Mack}}, \binits{J.}},
\bauthor{\bsnm{{Voit}}, \binits{G.M.}},
\bauthor{\bsnm{{Sparks}}, \binits{W.}},
\bauthor{\bsnm{{Elston}}, \binits{R.}},
\bauthor{\bsnm{{Maloney}}, \binits{P.R.}}:
\bjtitle{\apj}
\bvolume{545},
\bfpage{670}
(\byear{2000}).
\arxivurl{astro-ph/0007062}.
doi:\doiurl{10.1086/317836}
\end{barticle}
\endbibitem

\bibitem[\protect\citeauthoryear{{Draine}}{2003}]{Draine2003}
\begin{barticle}
\bauthor{\bsnm{{Draine}}, \binits{B.T.}}:
\bjtitle{\araa}
\bvolume{41},
\bfpage{241}
(\byear{2003}).
\arxivurl{astro-ph/0304489}
\end{barticle}
\endbibitem

\bibitem[\protect\citeauthoryear{{Draine}}{2011}]{Draine2011}
\begin{bbook}
\bauthor{\bsnm{{Draine}}, \binits{B.T.}}:
\bbtitle{Physics of the Interstellar and Intergalactic Medium},
(\byear{2011})
\end{bbook}
\endbibitem

\bibitem[\protect\citeauthoryear{{Draine} and {Li}}{2007}]{Draine2007}
\begin{barticle}
\bauthor{\bsnm{{Draine}}, \binits{B.T.}},
\bauthor{\bsnm{{Li}}, \binits{A.}}:
\bjtitle{\apj}
\bvolume{657},
\bfpage{810}
(\byear{2007}).
\arxivurl{astro-ph/0608003}.
doi:\doiurl{10.1086/511055}
\end{barticle}
\endbibitem

\bibitem[\protect\citeauthoryear{{Dwek} and {Krennrich}}{2005}]{Dwek2005}
\begin{barticle}
\bauthor{\bsnm{{Dwek}}, \binits{E.}},
\bauthor{\bsnm{{Krennrich}}, \binits{F.}}:
\bjtitle{\apj}
\bvolume{618},
\bfpage{657}
(\byear{2005}).
\arxivurl{astro-ph/0406565}.
doi:\doiurl{10.1086/426010}
\end{barticle}
\endbibitem

\bibitem[\protect\citeauthoryear{{Dwek} and {Krennrich}}{2013}]{Dwek2013}
\begin{barticle}
\bauthor{\bsnm{{Dwek}}, \binits{E.}},
\bauthor{\bsnm{{Krennrich}}, \binits{F.}}:
\bjtitle{Astroparticle Physics}
\bvolume{43},
\bfpage{112}
(\byear{2013}).
doi:\doiurl{10.1016/j.astropartphys.2012.09.003}
\end{barticle}
\endbibitem

\bibitem[\protect\citeauthoryear{{Finkelman} et~al.}{2008}]{Finkelman2008}
\begin{barticle}
\bauthor{\bsnm{{Finkelman}}, \binits{I.}},
\bauthor{\bsnm{{Brosch}}, \binits{N.}},
\bauthor{\bsnm{{Kniazev}}, \binits{A.Y.}},
\bauthor{\bsnm{{Buckley}}, \binits{D.A.H.}},
\bauthor{\bsnm{{O'Donoghue}}, \binits{D.}},
\bauthor{\bsnm{{Hashimoto}}, \binits{Y.}},
\bauthor{\bsnm{{Loaring}}, \binits{N.}},
\bauthor{\bsnm{{Romero-Colmenero}}, \binits{E.}},
\bauthor{\bsnm{{Still}}, \binits{M.}},
\bauthor{\bsnm{{Sefako}}, \binits{R.}},
\bauthor{\bsnm{{V{\"a}is{\"a}nen}}, \binits{P.}}:
\bjtitle{\mnras}
\bvolume{390},
\bfpage{969}
(\byear{2008}).
\arxivurl{0808.0714}.
doi:\doiurl{10.1111/j.1365-2966.2008.13785.x}
\end{barticle}
\endbibitem

\bibitem[\protect\citeauthoryear{{Finkelman} et~al.}{2010}]{Finkelman2010}
\begin{barticle}
\bauthor{\bsnm{{Finkelman}}, \binits{I.}},
\bauthor{\bsnm{{Brosch}}, \binits{N.}},
\bauthor{\bsnm{{Kniazev}}, \binits{A.Y.}},
\bauthor{\bsnm{{V{\"a}is{\"a}nen}}, \binits{P.}},
\bauthor{\bsnm{{Buckley}}, \binits{D.A.H.}},
\bauthor{\bsnm{{O'Donoghue}}, \binits{D.}},
\bauthor{\bsnm{{Gulbis}}, \binits{A.}},
\bauthor{\bsnm{{Hashimoto}}, \binits{Y.}},
\bauthor{\bsnm{{Loaring}}, \binits{N.}},
\bauthor{\bsnm{{Romero-Colmenero}}, \binits{E.}},
\bauthor{\bsnm{{Sefako}}, \binits{R.}}:
\bjtitle{\mnras}
\bvolume{409},
\bfpage{727}
(\byear{2010}).
\arxivurl{1008.5149}.
doi:\doiurl{10.1111/j.1365-2966.2010.17334.x}
\end{barticle}
\endbibitem

\bibitem[\protect\citeauthoryear{{Geller} et~al.}{1997}]{Geller1997}
\begin{barticle}
\bauthor{\bsnm{{Geller}}, \binits{M.J.}},
\bauthor{\bsnm{{Kurtz}}, \binits{M.J.}},
\bauthor{\bsnm{{Wegner}}, \binits{G.}},
\bauthor{\bsnm{{Thorstensen}}, \binits{J.R.}},
\bauthor{\bsnm{{Fabricant}}, \binits{D.G.}},
\bauthor{\bsnm{{Marzke}}, \binits{R.O.}},
\bauthor{\bsnm{{Huchra}}, \binits{J.P.}},
\bauthor{\bsnm{{Schild}}, \binits{R.E.}},
\bauthor{\bsnm{{Falco}}, \binits{E.E.}}:
\bjtitle{\aj}
\bvolume{114},
\bfpage{2205}
(\byear{1997}).
\arxivurl{astro-ph/9710109}.
doi:\doiurl{10.1086/118640}
\end{barticle}
\endbibitem

\bibitem[\protect\citeauthoryear{{Gilmore} et~al.}{2012}]{Gilmore2012}
\begin{barticle}
\bauthor{\bsnm{{Gilmore}}, \binits{R.C.}},
\bauthor{\bsnm{{Somerville}}, \binits{R.S.}},
\bauthor{\bsnm{{Primack}}, \binits{J.R.}},
\bauthor{\bsnm{{Dom{\'{\i}}nguez}}, \binits{A.}}:
\bjtitle{\mnras}
\bvolume{422},
\bfpage{3189}
(\byear{2012}).
\arxivurl{1104.0671}.
doi:\doiurl{10.1111/j.1365-2966.2012.20841.x}
\end{barticle}
\endbibitem

\bibitem[\protect\citeauthoryear{{Gonz{\'a}lez} et~al.}{1998}]{Gonzalez1998}
\begin{barticle}
\bauthor{\bsnm{{Gonz{\'a}lez}}, \binits{R.A.}},
\bauthor{\bsnm{{Allen}}, \binits{R.J.}},
\bauthor{\bsnm{{Dirsch}}, \binits{B.}},
\bauthor{\bsnm{{Ferguson}}, \binits{H.C.}},
\bauthor{\bsnm{{Calzetti}}, \binits{D.}},
\bauthor{\bsnm{{Panagia}}, \binits{N.}}:
\bjtitle{\apj}
\bvolume{506},
\bfpage{152}
(\byear{1998}).
\arxivurl{astro-ph/9806001}.
doi:\doiurl{10.1086/306242}
\end{barticle}
\endbibitem

\bibitem[\protect\citeauthoryear{{Goudfrooij} and {de
  Jong}}{1995}]{Goudfrooij1995}
\begin{barticle}
\bauthor{\bsnm{{Goudfrooij}}, \binits{P.}},
\bauthor{\bsnm{{de Jong}}, \binits{T.}}:
\bjtitle{\aap}
\bvolume{298},
\bfpage{784}
(\byear{1995}).
\arxivurl{astro-ph/9504011}
\end{barticle}
\endbibitem

\bibitem[\protect\citeauthoryear{{Goudfrooij} et~al.}{1994}]{Goudfrooij1994}
\begin{barticle}
\bauthor{\bsnm{{Goudfrooij}}, \binits{P.}},
\bauthor{\bsnm{{de Jong}}, \binits{T.}},
\bauthor{\bsnm{{Hansen}}, \binits{L.}},
\bauthor{\bsnm{{Norgaard-Nielsen}}, \binits{H.U.}}:
\bjtitle{\mnras}
\bvolume{271},
\bfpage{833}
(\byear{1994})
\end{barticle}
\endbibitem

\bibitem[\protect\citeauthoryear{{Graham} and {Driver}}{2005}]{Graham2005}
\begin{barticle}
\bauthor{\bsnm{{Graham}}, \binits{A.W.}},
\bauthor{\bsnm{{Driver}}, \binits{S.P.}}:
\bjtitle{\pasa}
\bvolume{22},
\bfpage{118}
(\byear{2005}).
\arxivurl{astro-ph/0503176}.
doi:\doiurl{10.1071/AS05001}
\end{barticle}
\endbibitem

\bibitem[\protect\citeauthoryear{{Harrison}}{1984}]{Harrison1984}
\begin{barticle}
\bauthor{\bsnm{{Harrison}}, \binits{E.R.}}:
\bjtitle{Science}
\bvolume{226},
\bfpage{941}
(\byear{1984})
\end{barticle}
\endbibitem

\bibitem[\protect\citeauthoryear{{Harrison}}{1990}]{Harrison1990}
\begin{bchapter}
\bauthor{\bsnm{{Harrison}}, \binits{E.R.}}:
In: \beditor{\bsnm{{Bowyer}}, \binits{S.}},
\beditor{\bsnm{{Leinert}}, \binits{C.}} (eds.)
\bbtitle{The Galactic and Extragalactic Background Radiation}.
\bsertitle{IAU Symposium},
vol. \bseriesno{139},
p. \bfpage{3}
(\byear{1990})
\end{bchapter}
\endbibitem

\bibitem[\protect\citeauthoryear{{Hauser} and {Dwek}}{2001}]{Hauser2001}
\begin{barticle}
\bauthor{\bsnm{{Hauser}}, \binits{M.G.}},
\bauthor{\bsnm{{Dwek}}, \binits{E.}}:
\bjtitle{\araa}
\bvolume{39},
\bfpage{249}
(\byear{2001}).
doi:\doiurl{10.1146/annurev.astro.39.1.249}
\end{barticle}
\endbibitem

\bibitem[\protect\citeauthoryear{{Holwerda} et~al.}{2005a}]{Holwerda2005b}
\begin{barticle}
\bauthor{\bsnm{{Holwerda}}, \binits{B.W.}},
\bauthor{\bsnm{{Gonzalez}}, \binits{R.A.}},
\bauthor{\bsnm{{Allen}}, \binits{R.J.}},
\bauthor{\bsnm{{van der Kruit}}, \binits{P.C.}}:
\bjtitle{\aj}
\bvolume{129},
\bfpage{1381}
(\byear{2005}a).
\arxivurl{astro-ph/0411662}.
doi:\doiurl{10.1086/427711}
\end{barticle}
\endbibitem

\bibitem[\protect\citeauthoryear{{Holwerda} et~al.}{2005b}]{Holwerda2005a}
\begin{barticle}
\bauthor{\bsnm{{Holwerda}}, \binits{B.W.}},
\bauthor{\bsnm{{Gonzalez}}, \binits{R.A.}},
\bauthor{\bsnm{{Allen}}, \binits{R.J.}},
\bauthor{\bsnm{{van der Kruit}}, \binits{P.C.}}:
\bjtitle{\aj}
\bvolume{129},
\bfpage{1396}
(\byear{2005}b).
\arxivurl{astro-ph/0411663}.
doi:\doiurl{10.1086/427716}
\end{barticle}
\endbibitem

\bibitem[\protect\citeauthoryear{{Holwerda} et~al.}{2007}]{Holwerda2007}
\begin{barticle}
\bauthor{\bsnm{{Holwerda}}, \binits{B.W.}},
\bauthor{\bsnm{{Draine}}, \binits{B.}},
\bauthor{\bsnm{{Gordon}}, \binits{K.D.}},
\bauthor{\bsnm{{Gonz{\'a}lez}}, \binits{R.A.}},
\bauthor{\bsnm{{Calzetti}}, \binits{D.}},
\bauthor{\bsnm{{Thornley}}, \binits{M.}},
\bauthor{\bsnm{{Buckalew}}, \binits{B.}},
\bauthor{\bsnm{{Allen}}, \binits{R.J.}},
\bauthor{\bsnm{{van der Kruit}}, \binits{P.C.}}:
\bjtitle{\aj}
\bvolume{134},
\bfpage{2226}
(\byear{2007}).
\arxivurl{0707.4165}.
doi:\doiurl{10.1086/522230}
\end{barticle}
\endbibitem

\bibitem[\protect\citeauthoryear{{Jones} et~al.}{2004}]{Jones2004}
\begin{barticle}
\bauthor{\bsnm{{Jones}}, \binits{B.J.}},
\bauthor{\bsnm{{Mart{\'{\i}}nez}}, \binits{V.J.}},
\bauthor{\bsnm{{Saar}}, \binits{E.}},
\bauthor{\bsnm{{Trimble}}, \binits{V.}}:
\bjtitle{Reviews of Modern Physics}
\bvolume{76},
\bfpage{1211}
(\byear{2004}).
\arxivurl{astro-ph/0406086}.
doi:\doiurl{10.1103/RevModPhys.76.1211}
\end{barticle}
\endbibitem

\bibitem[\protect\citeauthoryear{{Kneiske} et~al.}{2004}]{Kneiske2004}
\begin{barticle}
\bauthor{\bsnm{{Kneiske}}, \binits{T.M.}},
\bauthor{\bsnm{{Bretz}}, \binits{T.}},
\bauthor{\bsnm{{Mannheim}}, \binits{K.}},
\bauthor{\bsnm{{Hartmann}}, \binits{D.H.}}:
\bjtitle{\aap}
\bvolume{413},
\bfpage{807}
(\byear{2004}).
\arxivurl{astro-ph/0309141}.
doi:\doiurl{10.1051/0004-6361:20031542}
\end{barticle}
\endbibitem

\bibitem[\protect\citeauthoryear{{Knutsen}}{1997}]{Knutsen1997}
\begin{barticle}
\bauthor{\bsnm{{Knutsen}}, \binits{H.}}:
\bjtitle{European Journal of Physics}
\bvolume{18},
\bfpage{295}
(\byear{1997}).
doi:\doiurl{10.1088/0143-0807/18/4/010}
\end{barticle}
\endbibitem

\bibitem[\protect\citeauthoryear{{Koppen} and {Vergely}}{1998}]{Koppen1998}
\begin{barticle}
\bauthor{\bsnm{{Koppen}}, \binits{J.}},
\bauthor{\bsnm{{Vergely}}, \binits{J.-L.}}:
\bjtitle{\mnras}
\bvolume{299},
\bfpage{567}
(\byear{1998}).
doi:\doiurl{10.1046/j.1365-8711.1998.01808.x}
\end{barticle}
\endbibitem

\bibitem[\protect\citeauthoryear{{Lagache} et~al.}{2005}]{Lagache2005}
\begin{barticle}
\bauthor{\bsnm{{Lagache}}, \binits{G.}},
\bauthor{\bsnm{{Puget}}, \binits{J.-L.}},
\bauthor{\bsnm{{Dole}}, \binits{H.}}:
\bjtitle{\araa}
\bvolume{43},
\bfpage{727}
(\byear{2005})
\end{barticle}
\endbibitem

\bibitem[\protect\citeauthoryear{{Lisenfeld} et~al.}{2008}]{Lisenfeld2008}
\begin{bchapter}
\bauthor{\bsnm{{Lisenfeld}}, \binits{U.}},
\bauthor{\bsnm{{Rela{\~n}o}}, \binits{M.}},
\bauthor{\bsnm{{V{\'{\i}}lchez}}, \binits{J.}},
\bauthor{\bsnm{{Battaner}}, \binits{E.}},
\bauthor{\bsnm{{Hermelo}}, \binits{I.}}:
In: \beditor{\bsnm{{Hunt}}, \binits{L.K.}},
\beditor{\bsnm{{Madden}}, \binits{S.C.}},
\beditor{\bsnm{{Schneider}}, \binits{R.}} (eds.)
\bbtitle{IAU Symposium}.
\bsertitle{IAU Symposium},
vol. \bseriesno{255},
p. \bfpage{260}
(\byear{2008}).
\arxivurl{0808.1953}.
doi:\doiurl{10.1017/S1743921308024915}
\end{bchapter}
\endbibitem

\bibitem[\protect\citeauthoryear{{Madau} and {Pozzetti}}{2000}]{Madau2000}
\begin{barticle}
\bauthor{\bsnm{{Madau}}, \binits{P.}},
\bauthor{\bsnm{{Pozzetti}}, \binits{L.}}:
\bjtitle{\mnras}
\bvolume{312},
\bfpage{9}
(\byear{2000}).
\arxivurl{astro-ph/9907315}.
doi:\doiurl{10.1046/j.1365-8711.2000.03268.x}
\end{barticle}
\endbibitem

\bibitem[\protect\citeauthoryear{{Margolis} and {Schramm}}{1977}]{Margolis1977}
\begin{barticle}
\bauthor{\bsnm{{Margolis}}, \binits{S.H.}},
\bauthor{\bsnm{{Schramm}}, \binits{D.N.}}:
\bjtitle{\apj}
\bvolume{214},
\bfpage{339}
(\byear{1977}).
doi:\doiurl{10.1086/155256}
\end{barticle}
\endbibitem

\bibitem[\protect\citeauthoryear{{Mathis}}{1990}]{Mathis1990}
\begin{barticle}
\bauthor{\bsnm{{Mathis}}, \binits{J.S.}}:
\bjtitle{\araa}
\bvolume{28},
\bfpage{37}
(\byear{1990}).
doi:\doiurl{10.1146/annurev.aa.28.090190.000345}
\end{barticle}
\endbibitem

\bibitem[\protect\citeauthoryear{{Matsumoto} et~al.}{2005}]{Matsumoto2005}
\begin{barticle}
\bauthor{\bsnm{{Matsumoto}}, \binits{T.}},
\bauthor{\bsnm{{Matsuura}}, \binits{S.}},
\bauthor{\bsnm{{Murakami}}, \binits{H.}},
\bauthor{\bsnm{{Tanaka}}, \binits{M.}},
\bauthor{\bsnm{{Freund}}, \binits{M.}},
\bauthor{\bsnm{{Lim}}, \binits{M.}},
\bauthor{\bsnm{{Cohen}}, \binits{M.}},
\bauthor{\bsnm{{Kawada}}, \binits{M.}},
\bauthor{\bsnm{{Noda}}, \binits{M.}}:
\bjtitle{\apj}
\bvolume{626},
\bfpage{31}
(\byear{2005}).
\arxivurl{astro-ph/0411593}.
doi:\doiurl{10.1086/429383}
\end{barticle}
\endbibitem

\bibitem[\protect\citeauthoryear{{M{\'e}nard} et~al.}{2010}]{Menard2010a}
\begin{barticle}
\bauthor{\bsnm{{M{\'e}nard}}, \binits{B.}},
\bauthor{\bsnm{{Scranton}}, \binits{R.}},
\bauthor{\bsnm{{Fukugita}}, \binits{M.}},
\bauthor{\bsnm{{Richards}}, \binits{G.}}:
\bjtitle{\mnras}
\bvolume{405},
\bfpage{1025}
(\byear{2010}).
\arxivurl{0902.4240}.
doi:\doiurl{10.1111/j.1365-2966.2010.16486.x}
\end{barticle}
\endbibitem

\bibitem[\protect\citeauthoryear{{Milne} and {Aller}}{1980}]{Milne1980}
\begin{barticle}
\bauthor{\bsnm{{Milne}}, \binits{D.K.}},
\bauthor{\bsnm{{Aller}}, \binits{L.H.}}:
\bjtitle{\aj}
\bvolume{85},
\bfpage{17}
(\byear{1980}).
doi:\doiurl{10.1086/112628}
\end{barticle}
\endbibitem

\bibitem[\protect\citeauthoryear{{Muller} et~al.}{2008}]{Muller2008}
\begin{barticle}
\bauthor{\bsnm{{Muller}}, \binits{S.}},
\bauthor{\bsnm{{Wu}}, \binits{S.-Y.}},
\bauthor{\bsnm{{Hsieh}}, \binits{B.-C.}},
\bauthor{\bsnm{{Gonz{\'a}lez}}, \binits{R.A.}},
\bauthor{\bsnm{{Loinard}}, \binits{L.}},
\bauthor{\bsnm{{Yee}}, \binits{H.K.C.}},
\bauthor{\bsnm{{Gladders}}, \binits{M.D.}}:
\bjtitle{\apj}
\bvolume{680},
\bfpage{975}
(\byear{2008}).
\arxivurl{0801.2613}.
doi:\doiurl{10.1086/529583}
\end{barticle}
\endbibitem

\bibitem[\protect\citeauthoryear{{Nickerson} and
  {Partridge}}{1971}]{Nickerson1971}
\begin{barticle}
\bauthor{\bsnm{{Nickerson}}, \binits{B.G.}},
\bauthor{\bsnm{{Partridge}}, \binits{R.B.}}:
\bjtitle{\apj}
\bvolume{169},
\bfpage{203}
(\byear{1971}).
doi:\doiurl{10.1086/151133}
\end{barticle}
\endbibitem

\bibitem[\protect\citeauthoryear{{Peacock}}{1999}]{Peacock1999}
\begin{bbook}
\bauthor{\bsnm{{Peacock}}, \binits{J.A.}}:
\bbtitle{Cosmological Physics},
(\byear{1999})
\end{bbook}
\endbibitem

\bibitem[\protect\citeauthoryear{{Peebles}}{1993}]{Peebles1993}
\begin{bbook}
\bauthor{\bsnm{{Peebles}}, \binits{P.J.E.}}:
\bbtitle{Principles of Physical Cosmology},
(\byear{1993})
\end{bbook}
\endbibitem

\bibitem[\protect\citeauthoryear{{Peebles}}{2001}]{Peebles2001}
\begin{barticle}
\bauthor{\bsnm{{Peebles}}, \binits{P.J.E.}}:
\bjtitle{\apj}
\bvolume{557},
\bfpage{495}
(\byear{2001}).
\arxivurl{astro-ph/0101127}.
doi:\doiurl{10.1086/322254}
\end{barticle}
\endbibitem

\bibitem[\protect\citeauthoryear{{Popescu} et~al.}{2011}]{Popescu2011}
\begin{barticle}
\bauthor{\bsnm{{Popescu}}, \binits{C.C.}},
\bauthor{\bsnm{{Tuffs}}, \binits{R.J.}},
\bauthor{\bsnm{{Dopita}}, \binits{M.A.}},
\bauthor{\bsnm{{Fischera}}, \binits{J.}},
\bauthor{\bsnm{{Kylafis}}, \binits{N.D.}},
\bauthor{\bsnm{{Madore}}, \binits{B.F.}}:
\bjtitle{\aap}
\bvolume{527},
\bfpage{109}
(\byear{2011}).
\arxivurl{1011.2942}.
doi:\doiurl{10.1051/0004-6361/201015217}
\end{barticle}
\endbibitem

\bibitem[\protect\citeauthoryear{{Primack} et~al.}{2011}]{Primack2011}
\begin{bchapter}
\bauthor{\bsnm{{Primack}}, \binits{J.R.}},
\bauthor{\bsnm{{Dom{\'{\i}}nguez}}, \binits{A.}},
\bauthor{\bsnm{{Gilmore}}, \binits{R.C.}},
\bauthor{\bsnm{{Somerville}}, \binits{R.S.}}:
In: \beditor{\bsnm{{Aharonian}}, \binits{F.A.}},
\beditor{\bsnm{{Hofmann}}, \binits{W.}},
\beditor{\bsnm{{Rieger}}, \binits{F.M.}} (eds.)
\bbtitle{American Institute of Physics Conference Series}.
\bsertitle{American Institute of Physics Conference Series},
vol. \bseriesno{1381},
p. \bfpage{72}
(\byear{2011}).
\arxivurl{1107.2566}.
doi:\doiurl{10.1063/1.3635825}
\end{bchapter}
\endbibitem

\bibitem[\protect\citeauthoryear{{Romani} and {Maoz}}{1992}]{Romani1992}
\begin{barticle}
\bauthor{\bsnm{{Romani}}, \binits{R.W.}},
\bauthor{\bsnm{{Maoz}}, \binits{D.}}:
\bjtitle{\apj}
\bvolume{386},
\bfpage{36}
(\byear{1992}).
doi:\doiurl{10.1086/170989}
\end{barticle}
\endbibitem

\bibitem[\protect\citeauthoryear{{Schechter}}{1976}]{Schechter1976}
\begin{barticle}
\bauthor{\bsnm{{Schechter}}, \binits{P.}}:
\bjtitle{\apj}
\bvolume{203},
\bfpage{297}
(\byear{1976})
\end{barticle}
\endbibitem

\bibitem[\protect\citeauthoryear{{Schlegel} et~al.}{1998}]{Schlegel1998}
\begin{barticle}
\bauthor{\bsnm{{Schlegel}}, \binits{D.J.}},
\bauthor{\bsnm{{Finkbeiner}}, \binits{D.P.}},
\bauthor{\bsnm{{Davis}}, \binits{M.}}:
\bjtitle{\apj}
\bvolume{500},
\bfpage{525}
(\byear{1998}).
\arxivurl{astro-ph/9710327}.
doi:\doiurl{10.1086/305772}
\end{barticle}
\endbibitem

\bibitem[\protect\citeauthoryear{{Shen} et~al.}{2003}]{Shen2003}
\begin{barticle}
\bauthor{\bsnm{{Shen}}, \binits{S.}},
\bauthor{\bsnm{{Mo}}, \binits{H.J.}},
\bauthor{\bsnm{{White}}, \binits{S.D.M.}},
\bauthor{\bsnm{{Blanton}}, \binits{M.R.}},
\bauthor{\bsnm{{Kauffmann}}, \binits{G.}},
\bauthor{\bsnm{{Voges}}, \binits{W.}},
\bauthor{\bsnm{{Brinkmann}}, \binits{J.}},
\bauthor{\bsnm{{Csabai}}, \binits{I.}}:
\bjtitle{\mnras}
\bvolume{343},
\bfpage{978}
(\byear{2003}).
doi:\doiurl{10.1046/j.1365-8711.2003.06740.x}
\end{barticle}
\endbibitem

\bibitem[\protect\citeauthoryear{{Somerville} et~al.}{2012}]{Somerville2012}
\begin{barticle}
\bauthor{\bsnm{{Somerville}}, \binits{R.S.}},
\bauthor{\bsnm{{Gilmore}}, \binits{R.C.}},
\bauthor{\bsnm{{Primack}}, \binits{J.R.}},
\bauthor{\bsnm{{Dom{\'{\i}}nguez}}, \binits{A.}}:
\bjtitle{\mnras}
\bvolume{423},
\bfpage{1992}
(\byear{2012}).
\arxivurl{1104.0669}.
doi:\doiurl{10.1111/j.1365-2966.2012.20490.x}
\end{barticle}
\endbibitem

\bibitem[\protect\citeauthoryear{{Tuffs} et~al.}{2004}]{Tuffs2004}
\begin{barticle}
\bauthor{\bsnm{{Tuffs}}, \binits{R.J.}},
\bauthor{\bsnm{{Popescu}}, \binits{C.C.}},
\bauthor{\bsnm{{V{\"o}lk}}, \binits{H.J.}},
\bauthor{\bsnm{{Kylafis}}, \binits{N.D.}},
\bauthor{\bsnm{{Dopita}}, \binits{M.A.}}:
\bjtitle{\aap}
\bvolume{419},
\bfpage{821}
(\byear{2004}).
\arxivurl{astro-ph/0401630}.
doi:\doiurl{10.1051/0004-6361:20035689}
\end{barticle}
\endbibitem

\bibitem[\protect\citeauthoryear{{Voit} and {Donahue}}{1997}]{Voit1997}
\begin{barticle}
\bauthor{\bsnm{{Voit}}, \binits{G.M.}},
\bauthor{\bsnm{{Donahue}}, \binits{M.}}:
\bjtitle{\apj}
\bvolume{486},
\bfpage{242}
(\byear{1997}).
\arxivurl{astro-ph/9706107}
\end{barticle}
\endbibitem

\bibitem[\protect\citeauthoryear{{von Benda-Beckmann} and
  {M{\"u}ller}}{2008}]{vonBenda_Beckmann2008}
\begin{barticle}
\bauthor{\bsnm{{von Benda-Beckmann}}, \binits{A.M.}},
\bauthor{\bsnm{{M{\"u}ller}}, \binits{V.}}:
\bjtitle{\mnras}
\bvolume{384},
\bfpage{1189}
(\byear{2008}).
doi:\doiurl{10.1111/j.1365-2966.2007.12789.x}
\end{barticle}
\endbibitem

\bibitem[\protect\citeauthoryear{{Wesson}}{1989}]{Wesson1989}
\begin{barticle}
\bauthor{\bsnm{{Wesson}}, \binits{P.S.}}:
\bjtitle{Journal of the British Astronomical Association}
\bvolume{99},
\bfpage{10}
(\byear{1989})
\end{barticle}
\endbibitem

\bibitem[\protect\citeauthoryear{{Wesson} et~al.}{1987}]{Wesson1987}
\begin{barticle}
\bauthor{\bsnm{{Wesson}}, \binits{P.S.}},
\bauthor{\bsnm{{Valle}}, \binits{K.}},
\bauthor{\bsnm{{Stabell}}, \binits{R.}}:
\bjtitle{\apj}
\bvolume{317},
\bfpage{601}
(\byear{1987}).
doi:\doiurl{10.1086/165306}
\end{barticle}
\endbibitem

\bibitem[\protect\citeauthoryear{{Xie} et~al.}{2015}]{Xie2015}
\begin{barticle}
\bauthor{\bsnm{{Xie}}, \binits{X.}},
\bauthor{\bsnm{{Shen}}, \binits{S.}},
\bauthor{\bsnm{{Shao}}, \binits{Z.}},
\bauthor{\bsnm{{Yin}}, \binits{J.}}:
\bjtitle{\apjl}
\bvolume{802},
\bfpage{16}
(\byear{2015}).
doi:\doiurl{10.1088/2041-8205/802/2/L16}
\end{barticle}
\endbibitem

\end{thebibliography}


\end{document}